\documentstyle[prd,aps,multicol,eqsecnum,amssymb]{revtex}
\begin{document}
\draft

\title{Supergravity interacting with bosonic p-branes and local supersymmetry}

\author{Igor A. Bandos$^{\ast,\ddagger}$ $^\star$, Jos\'e A. de 
Azc\'arraga$^{\ast}$ $^\star$ and 
Jos\'e M. Izquierdo$^{\dagger}$}
\address{
$^{\ast}$Departamento de F\'{\i}sica Te\'orica and IFIC 
(CSIC-UVEG), 
 46100-Burjassot (Valencia), Spain} 
\address{$^{\dagger}$Departamento de F\'{\i}sica Te\'orica,
 Facultad de Ciencias, 47011-Valladolid, Spain}
\address{$^{\ddagger}$Institute for Theoretical Physics, NSC KIPT, 
UA61108,
Kharkov, Ukraine} 
\address{ $^\star$ The Erwin 
Schr\"{o}dinger Institute for Mathematical Physics, Boltzmanngasse 9,
1090-Wien, Austria}

\date{FTUV/01-2112, IFIC/01-70; December 21, 2001}

\maketitle 

\def\theequation{\arabic{equation}}

\begin{abstract}

We study the coupling of supergravity with   
a purely bosonic brane source (bosonic $p$--brane). 
The  interaction, described by the sum of their respective  actions, 
is self-consistent if the bosonic $p$--brane is the pure
bosonic limit of a super--$p$--brane. 
In that case the dynamical system 
preserves $1/2$ of the local supersymmetry 
characteristic of the `free' supergravity.

\end{abstract}


\renewcommand{\theequation}{\arabic{section}.\arabic{equation}} 
\setcounter{equation}0

\begin{multicols}{2}

\narrowtext

\section{Introduction}

It is usually expected that the addition of a pure bosonic system to a 
supersymmetric one must produce a complete breaking of the supersymmetry. 
Nevertheless, it has been shown \cite{Bergshoeff} that 
the  coupled system of supergravity on the 
orbifold spacetime $M^{9}\times [S^1/\hbox{\bf Z}_2]$ 
and  {\sl pure bosonic} branes {\sl fixed} at the orbifold fixed  
`points' preserves $1/2$ 
of the local supersymmetry, while the other $1/2$ components of its parameter 
 vanish on the brane due to the orbifold projection  
(see  \cite{BKPO,Bagger} for $D=5$ models; most of the models 
\cite{BKPO,Bagger,Bergshoeff} are for domain walls, $p=D-2$).
These systems have been discussed in connection with the 
search for a supersymmetric 
generalization \cite{Duff} of the Randall-Sundrum  Brane World scenario 
\cite{RS}.

In this paper we show that, in general, a {\sl dynamical} bosonic brane 
source interacting with {\sl dynamical} supergravity preserves $1/2$ of the 
local supersymmetry $\delta_{ls}$ exhibited by the supergravity 
action if the bosonic $p$--brane can be regarded as the bosonic limit  
of a super--$p$--brane. This is still sufficient to   
preserve the selfconsistency of the gravitino interaction. 
Moreover, on the worldvolume the local supersymmetry parameter  
has the characteristic structure of the $\kappa$--symmetry 
transformation of the superbrane.  

Our notation is close to that in  \cite{BAIL}. 
The (non-scalar) physical fields of 
the $D$--dimensional supergravity multiplet 
(graviton $e_{\mu}^{~ {a}}(x)$, gravitino 
$\psi_{\mu}^{\underline{\alpha}}(x)$, antisymmetric rank $q$ gauge 
field(s) $ C_{ \mu_1 \ldots  \mu_q}(x)$ and the 
spin connection field $w_{\mu}^{ {a} {b}}(x)$) are given 
by differential forms on spacetime $M^D$
\begin{eqnarray}\label{sgmult}
& e^a(x)= d{x}^{\mu} e_{\mu}^{~ {a}}(x) \quad 
[a, \; \mu = 0, 1,\ldots , (D-1)]\; , 
\\ 
\label{sgmultg}
& C_{q}\equiv {1\over q!} dx^{ \mu_q}\wedge \ldots\wedge   
dx^{ \mu_1} C_{ \mu_1 \ldots  \mu_q}(x) \; ,
\\ \nonumber 
& e^{\underline{\alpha}}(x)= d{x}^{\mu} \psi_{\mu}^{\underline{\alpha}}(x)
= e^a \psi_{a}^{\underline{\alpha}}(x)
\;   
\\ \label{sgmultf}
&  [
\underline{\alpha} = 1,\ldots , 
n\; , \qquad n={\hbox{dim}}(Spin(1,D-1))] \; , \\ 
\label{sgmultw}
& w^{ {a} {b}}(x)= dx^{\mu} w_{\mu}^{ {a} {b}}(x)=-w^{ {b} {a}}(x)\; . \qquad 
\end{eqnarray}
For simplicity we will not address here the 
cases where the supergravity multiplet contains scalar fields 
as well as when the brane carries  worldvolume gauge fields. 
Thus our basic examples are 
$N$=$1$, $D$=$3,4, 11$ supergravity 
interacting with a bosonic $p$--brane.

\medskip  

\section{Action for the coupled system  
of supergravity and a bosonic brane source}

The coupled system action is given by the sum 
\begin{eqnarray}\label{SSG+p}
& S= S_{D,SG}+ S_{D,p,0}\equiv  \int_{M^{D}}  {{\cal L}}_{D}
+  \int_{W^{p+1}}  \hat{{\cal L}}_{p+1} \; ,    
\qquad 
\end{eqnarray}  
where $S_{D,SG}$ 
and $S_{D,p,0}$ 
are the supergravity and {\sl bosonic} (hence the subindex 0) 
$p$--brane 
 actions. ${\cal L}_D$ is the Lagrangian $D$-form on $M^D$, 
\begin{eqnarray}\label{LD1}
& {\cal L}_D = 
d^D x L_{sg}(x)= {\cal L}^{2}_D + {\cal L}^{3/2}_D + {\cal L}_D{}^{\leq 1}\; , 
\end{eqnarray}
and ${\hat{\cal L}}_{p+1}$ is the Lagrangian $(p+1)$--form 
 on the worldvolume $W^{p+1}\subset M^D$ with local coordinates 
$\xi^i = (\tau, \sigma^1, \ldots , \sigma^p)$, 
\begin{eqnarray}\label{LpST}
& {\hat{\cal L}}_{p+1}= d^{p+1} \xi\, L_{brane}(\xi)  = 
{1\over 2(p+1)} * \hat{e}_{{a}} \wedge \hat{e}^{{a}} - \hat{C}_{p+1}\; 
\end{eqnarray}
({\it cf.} \cite{BST,AETW} and refs. therein); the hat will be used from
now on to denote the dependence on $\xi\in W^{p+1}$. 
The forms on $W^{p+1}$ in Eq. (\ref{LpST}),   
\begin{eqnarray}\label{hEa}
&  \hat{e}^{ {a}} = d\hat{x}^{\mu}(\xi ) e_{\mu}^{~ {a}}(\hat{x})
= d\xi^i \partial_i \hat{x}^{\mu}(\xi )  e_{\mu }^{~ {a}}(\hat{x})\; , 
\\    \label{hCp}
& \hat{C}_{p+1}= {1\over (p+1)!} d\hat{x}^{ \mu_{p+1}}\ldots  
d\hat{x}^{ \mu_1} C_{ \mu_1 \ldots  \mu_{p+1}}(\hat{x})\, ,
\end{eqnarray}  
are the pull--backs 
$\hat{\phi}^*(e^a)$, $\hat{\phi}^*(C_{p+1})$
of the vielbein (\ref{sgmult}) and 
gauge field (\ref{sgmultg}) forms on spacetime by the map 
\begin{eqnarray}\label{Wp1}
& 
\hat{\phi}: W^{p+1}\rightarrow M^{D}\, , 
\qquad  \hat{\phi}: \xi^i \mapsto 
\hat{x}^{ {\mu}}(\xi)\; . 
\qquad  
\end{eqnarray} 

When  $p\not= q-1$, {\it i.e.} if there is no 
$C_{p+1}$ in the supergravity 
multiplet (Eq. (\ref{sgmultg})), then $\hat{C}$ is also 
absent in (\ref{LpST})  and 
 ${\hat{\cal L}}_{p+1}$ reduces to the Nambu--Goto term 

\begin{eqnarray}\label{NGac}
{1\over 2(p+1)}\; * \hat{e}_{{a}} \wedge \hat{e}^{ {a}}=   
{1\over 2}\; d^{p+1}\xi \; \sqrt{|g|}\; , 
\end{eqnarray} 
where 
 $*$ is the Hodge operator for a ($p$+$1$)--dimensional 
space  with induced  
worldvolume metric $g_{ij}\equiv \hat{e}_i^a \hat{e}_{ja}$ on $W^{p+1}$. 
For $p$=$0$ (massless bosonic particle)  
and, {\it e.g.}, $D$=$4, 11$, for which  there is no $C_1$ in the 
supergravity multiplet, Eq. 
(\ref{LpST}) reduces to 
\begin{eqnarray}\label{L1}
& \hat{{\cal L}}_1 = 
{1 \over 2} l(\tau ) \hat{e}^a 
  \hat{e}^b_{\tau}\eta_{ab}\; , \qquad {} \qquad \\ \nonumber & 
\qquad \hat{e}^a=  d\tau \partial_\tau \, \hat{x}^\mu(\tau) \, 
e_\mu^a(\hat{x})\, 
\equiv d\tau \hat{e}^a_{\tau}(\tau)\; , \quad 
\end{eqnarray}
where $l(\tau )$ is 
a Lagrange multiplier 
(or worldline einbein) and the star operator can be defined by 
$*\hat{e}_a : = l(\tau)\hat{e}_{\tau a}\;$.  
Note that, as $\delta \sqrt{|g|}=   \sqrt{|g|} 
g^{ij}\hat{e}_{ia} \delta \hat{e}_j^a$, 
\begin{eqnarray}\label{vLpST}
& \delta {\hat{\cal L}}_{p+1}=  
{1\over2} * \hat{e}_{{a}} \wedge \delta \hat{e}^{{a}} 
- \delta \hat{C}_{p+1}\; , \\ 
\label{vL0ST}
& \delta {\hat{\cal L}}_{1}= l(\tau ) 
  \hat{e}_{\tau a}\; 
\delta \hat{e}^a + 
\; d\tau {1 \over 2} \delta l(\tau ) \hat{e}_\tau^a 
  \hat{e}_{\tau a}\; . 
\end{eqnarray}

\medskip

\subsection{Pure supergravity action and equations of motion}

The most important terms 
in the {\sl  supergravity Lagrangian form} (\ref{LD1}) are 
\begin{eqnarray}
\label{LD2}
& {\cal L}^{2}_D = R^{ {a} {b}} \wedge e^{\wedge (D-2)}_{ {a} {b}} \; , 
\\ 
\label{LD3/2}
& {\cal L}^{3/2}_D = 
-{2i\over 3} {\cal D} e^{\underline{\alpha}} \wedge e^{\underline{\beta}}
\wedge 
e^{\wedge (D-3)}_{ {a} {b} {c}}~
{\Gamma}^{ {a} {b} {c}}_{\underline{\alpha}\underline{\beta}} \; .
\end{eqnarray}
Here 
$ e^{\wedge (D-q)}_{ {a}_1\ldots  {a}_q}  
 \equiv 
{1 \over (D-q)!} 
\varepsilon_{ {a}_1\ldots  {a}_q {b}_1\ldots  {b}_{D-q}}
e^{ {b}_1} \wedge \ldots 
 \wedge e^{ {b}_{D-q}} \; $,
\begin{eqnarray}
\label{Rab1} 
& R^{ {a} {b}}= dw^{ {a} {b}}- w^{ {a} {c}}\wedge
w_{ {c}}^{~ {b}} \; 
\end{eqnarray} 
is the curvature 
of the spin connection  (\ref{sgmultw}), and 
\begin{eqnarray}
\label{Tal1}
& {\cal D}{e}^{\underline{\alpha}} ={T}^{\underline{\alpha}}=   
d{e}^{\underline{\alpha}}- {e}^{\underline{\beta}} 
\wedge  {w}_{\underline{\beta}}^{~\underline{\alpha}}\; , 
\end{eqnarray}
where $ {w}_{\underline{\beta}}^{~\underline{\alpha}}:=
{1\over 4} \; w^{ {a} {b}} 
\; \Gamma_{ {a} {b}}{}_{\underline{\beta}}^{~\underline{\alpha}}$,  
is the gravitino field strength.

We prefer here the first order 
formalism where the spin connection (\ref{sgmultw}) 
is considered as an independent variable and  
the equations of motion 
$\delta S_{D,SG} /\delta {w}^{ab}=0$ determine  
the `improved'  constraint on the spacetime torsion $T^a$,  
\begin{eqnarray}
& \label{Ta}  
T^a + i e^{\underline{\alpha}} \wedge e^{\underline{\beta}} 
\Gamma^a_{\underline{\alpha}\underline{\beta}}=0 
\quad \\ \nonumber & \qquad  ({T}^{ {a}} := {\cal D}e^a= 
 d{e}^{ {a}}-  {e}^{ {b}} \wedge  
 {w}_{ {b}}^{~ {a}})\; ,  
\end{eqnarray}
which expresses the spin connection through 
$e_\mu^a(x)$ and $\psi_\mu^{\underline{\alpha}}(x)$. 

In low dimensions, $D=3,4$, the Lagrangian form contains only 
the two terms (\ref{LD2}), (\ref{LD3/2}).
In higher dimensions ${\cal L}_D$ also includes, in particular, 
the kinetic term for the gauge fields (\ref{sgmultg}). 
This term can also 
be written in first order form. 
To this end one 
introduces  the auxiliary $(q+1)$--form
(see \cite{rheo}) 
\begin{eqnarray}
\label{Fp2} 
& F_{q+1} \equiv 
{1\over (q+1)!} e^{a_{q+1}}\wedge \ldots \wedge e^{a_1} \,  
F_{a_1\ldots a_{q+1}}(x)
\; , 
\end{eqnarray}
where $F^{a_1\ldots a_{q+1}}$ 
can be used as well to build the bosonic $(D-q-1)$--form 
\begin{eqnarray}
\label{FDp2} 
& {\cal F}_{D-q-1} = e^{\wedge (D-q-1)}_{a_1\ldots a_{q+1}} \,  
F^{a_1\ldots a_{q+1}}(x)\; . 
\end{eqnarray}
Then, ${\cal L}_D{}^{\leq 1}$ in Eq. (\ref{LD1}) is 
\begin{eqnarray}\label{LDC}
& {\cal L}_D{}^{\leq 1} = c
({\cal H}_{q+1} - {1\over 2} F_{q+1}) \wedge 
{\cal F}_{D-q-1} + \ldots \; , 
\\ \label{Hp+2}
&  {\cal H}_{q+1} := 
d{C}_{q} - 
c_1 {e}^{\underline{\alpha}} \wedge 
{e}^{\underline{\epsilon}}  \wedge  
\bar{\Gamma}^{(q-1)}{}_{\underline{\alpha}\underline{\epsilon}} \; , 
\\ \label{bG(k)}
& \bar{\Gamma}^{(k)}{}_{\underline{\alpha}\underline{\epsilon}}:= 
{1\over k!} {e}^{a_1} \wedge \ldots \wedge  
{e}^{a_k} 
\Gamma_{a_1\ldots a_k} {}_{\underline{\alpha}\underline{\epsilon}}\; , 
\end{eqnarray}
where $c, c_1$ are constants depending on 
$D$ and $q$,    ${\cal H}_{q+1}$  is the generalized 
field strength of ${C}_{q}$ and the terms denoted by dots do not contain 
$F^{a_1\ldots a_{q+1}}(x)$. 
The variations  $\delta {F}^{a_1 ... a_{q+1}}$ and 
$\delta {C}_{p+1}$ of the `free' supergravity action 
$ S_{D,SG}=  \int_{M^{D}}
({\cal L}^{2}_D + {\cal L}^{3/2}_D + {\cal L}_D{}^{\leq 1})$
produce
the first--order form of the free gauge field equations 
\begin{eqnarray}
\label{H} 
& {\cal H}_{q+1} -  F_{q+1} = 0\; \, ,  \\ 
\label{G}
&{G}_{(D-q)}\equiv 
d ({e}^{\wedge (D-q-1)}_{a_1\ldots a_{q+1}} F^{a_1\ldots a_{q+1}})
+ \ldots=0 \; , 
\end{eqnarray}
and $\delta {e}^{\underline{\alpha}}$ and $\delta {e}^{a}$ provide the 
Rarita--Schwinger and Einstein equations   
\begin{eqnarray} 
\label{RS}
& {{\Psi}}_{(D-1)\underline{\alpha}}
:= {{4i\over 3}}{\cal D} 
{e}^{\underline{\epsilon}} \wedge  
{e}^{\wedge (D-3)}_{ {a} {b} {c}}~
{\Gamma}^{ {a} {b} {c}}_{\underline{\epsilon}\underline{\alpha}} 
+\ldots = 0\; , \\  
\label{MD-1}
& M_{(D-1)~{ {a}}}:= {R}^{bc} \wedge  
{e}^{\wedge (D-3)}_{abc} + 
\ldots =0 \; .  
\end{eqnarray}

In the above notation
a generic variation of the supergravity action reads 
\begin{eqnarray}\label{vSSGD} 
\nonumber 
& \hspace{-0.6cm} \delta S_{D,SG}= - (-1)^D \int_{M^D}
M_{(D-1)a}\wedge \delta e^a + \qquad \hspace{-0.6cm} \\ 
\nonumber & + 
 (-1)^D \int_{M^D} \Psi_{(D-1)\underline{\alpha}} \wedge 
\delta e^{\underline{\alpha}} + \hspace{-0.6cm} \\ 
& + {(-1)^D} \int_{M^D} e^{\wedge (D-3)}_{abc} \wedge 
(T^c + i e^{\underline{\gamma}}\wedge e^{\underline{\delta}} 
\Gamma^c_{\underline{\gamma}\underline{\delta}}) \wedge \delta w^{ab} + 
\nonumber \hspace{-0.6cm}  \\ \nonumber & 
+ c \int_{M^D}({\cal H}_{q+1}-F_{q+1}) \wedge 
e^{\wedge (D-q-1)}_{a_1\ldots a_{q+1}} \;  \delta F^{a_1\ldots a_{q+1}} + 
\hspace{-0.6cm} 
\\ &  \qquad + c(-1)^{Dp}\int_{M^D} G_{D-q} \wedge  \delta C_q \; .
\hspace{-0.6cm} 
\end{eqnarray}

We note for future use that {\it e.g.}, the $(D-1)$--form $M_{(D-1)a}$ above 
is related with the usual expression for the equations of motion 
$\delta S/\delta e_\mu^a(x)$ by 
\begin{eqnarray}
\label{Mcomp}
& -(-1)^D M_{(D-1)a}= 
(dx)^{\wedge (D-1)}_\mu {\delta S \over \delta e_\mu^a(x)}
\\ \nonumber 
& \qquad \equiv -(-1)^D (dx)^{\wedge (D-1)}_\mu M_a^\mu\; , 
\end{eqnarray}
where $(dx)^{\wedge (D-1)}_\mu \wedge dx^\nu= d^Dx \, \delta_\mu^\nu$. 
Also, its external covariant derivative  
${\cal D}M_{(D-1)a}$ is the $D$--form 
\begin{eqnarray}
\label{DMcomp}
& {\cal D}M_{(D-1)a}= (dx)^{\wedge (D-1)}_\mu \wedge dx^\nu{\cal D}_\nu
 M_a^\mu  \nonumber \\ 
& \; =  d^Dx 
{\cal D}_\mu {\delta S \over \delta e_\mu^a(x)}\; .
\end{eqnarray}

\subsection{Local supersymmetry  of the pure supergravity action 
and  symmetry under spacetime general coordinate 
transformations} 

The `free' supergravity action 
$S_{D,SG}$ possesses {\sl local supersymmetry} $\delta_{ls}$  
(see \cite{rheo,rheo1}),  
\begin{eqnarray}
\label{lsusyx} & \delta_{ls}x^{\mu} =0\; ,  \\ 
\label{lsusy}
& \delta_{ls} e^a = -2i e^{\underline{\alpha}} 
\Gamma^a_{\underline{\alpha}\underline{\epsilon}}
\epsilon^{\underline{\epsilon}}(x)\; ,\quad  
\\  \label{lsusyf} 
& \delta_{ls}e^{\underline{\alpha}} = 
{\cal D} {\epsilon}^{\underline{\alpha}}(x)+ 
{\epsilon}^{\underline{\epsilon}}(x) 
{\cal M}_1{}_{\underline{\epsilon}}{}^{\underline{\alpha}}
 \; ,\quad \\  
\label{lsusyC}  
& \delta_{ls} C_{p+1} = 2c_1  {e}^{\underline{\alpha}} 
\wedge 
\bar{\Gamma}^{(p)}{}_{\underline{\alpha}\underline{\epsilon}}\; 
{\epsilon}^{\underline{\epsilon}}(x) \; , 
\\  \label{lsusyw}
& \delta_{ls} {w}^{ab}= 
{\cal W}_{1\underline{\epsilon}}^{ab} {\epsilon}^{\underline{\epsilon}}(x) 
 \, ,\;  
  \delta_{ls} {F}^{a_1 ... a_{q+1}} = 
S^{a_1 ... a_{q+1}}_{\underline{\epsilon}}
{\epsilon}^{\underline{\epsilon}} (x)\, ,
\end{eqnarray}
where ${\epsilon}^{\underline{\alpha}} (x)$ 
is the local supersymmetry parameter, 
$S^{a_1 ... a_{q+1}}_{\underline{\epsilon}}(x)$ and the one--forms 
${\cal M}_1{}_{\underline{\epsilon}}{}^{\underline{\alpha}}$, 
${\cal W}_1{}_{\underline{\epsilon}}{}^{ab}$ 
are constructed from the fields of the 
supergravity multiplet 
(\ref{sgmult})--(\ref{sgmultw})
 and the auxiliary fields ${F}_{a_1 ... a_{p+2}}(x)$.  

Besides the local supersymmetry 
(\ref{lsusyx})-- (\ref{lsusyw}), the action is invariant 
under  {\sl spacetime general coordinate transformations} 
$\delta_{gc}$ (see Appendix A) parametrized by $t^\mu (x)$,  
\begin{eqnarray}\label{gc}
& \delta_{gc}x^\mu & := x^{\mu \prime} - x^\mu= 
t^\mu (x) \; , \qquad   \\ \label{gce1}
& \delta_{gc} e^a & := e^a(x^\prime)- e^a(x) = \qquad 
\\ \nonumber && = 
{\cal D}t^a(x) + e^b t^c T_{cb}{}^a + e^b i_t w_b{}^{a}\; , \qquad etc. 
\end{eqnarray}
If the local Lorentz invariance of the theory is used, we can define 
\begin{eqnarray}
& \delta_{gc} e^a = {\cal D}t^a + e^b t^c T_{cb}{}^a \; , \qquad etc. \; ;   
\end{eqnarray}
see Appendix A 
and Eqs. 
(\ref{gc001})--
(\ref{gcF1}) for the transformations 
of all the supermultiplet components.

Note that it is also practical (and equivalent; see Appendix A) 
to define the action of 
$\delta_{gc}$ by its `variational version' (see  \cite{WZ})  
$\tilde{\delta}_{gc}$,  
\begin{eqnarray}\label{lt}
& \tilde{\delta}_{gc}x^\mu =0\; ,\quad   
\tilde{\delta}_{gc} e_\mu^a := {\cal D}_\mu t^a(x) + e_\mu^b t^c T_{cb}{}^a
 \; , \quad  etc. \; ,
\end{eqnarray}
{\it i.e.} by  {\it identifying} $\,\tilde{\delta}_{gc} e_\mu^a$, 
Eq. (\ref{gcvc}), with  
the components $({\delta}_{gc} e^a)_\mu$
of the differential form 
${\delta}_{gc} e^a$ in (\ref{gc}) 
without varying the spacetime coordinates $x^\mu$. 
The possibility of repacing  $\delta_{gc}$ in (\ref{gc})
by its equivalent `variational version'  
$\tilde{\delta}_{gc}$ in (\ref{lt})
is associated with the independence of differential 
forms on the choice of local coordinates 
{\it i.e.}, with the diffeomorphism invariance of  differential forms, 
$e^{\prime a}(x^{\prime})= e^{a}(x)$ {\it etc}. 
We shall come back to this point in connection with the second Noether 
theorem below.

\subsection{Local symmetries and Noether identities for the pure supergravity 
action}

In the framework of the second Noether theorem,  
the local supersymmetry (\ref{lsusyx})-- (\ref{lsusyw})
and the general coordinate transformation symmetry 
$\tilde{\delta}_{gc}$ (\ref{lt}) 
are reflected, respectively, by the (Noether) identities (see \cite{BAIL})  
\begin{eqnarray}\label{DRS}
& {\cal N}_{(D-1)\underline{\alpha}}:= &   
{\cal D} {\Psi}_{(D-1)\underline{\alpha}} -  \nonumber \\ 
 & & - 
2i {M}_{(D-1) a}\wedge {e}^{\underline{\epsilon}}
\Gamma^a_{\underline{\alpha}\underline{\epsilon}} + \ldots \equiv 0\; ,  
\\ 
\label{DMa=}
& {\cal N}_{(D-1)a}:= & {\cal D} {M}_{(D-1) a}- \ldots \equiv 0 \; , 
\end{eqnarray}
where the terms in the  $(D-1)$--forms 
${\cal N}_{(D-1)\underline{\alpha}}$ and ${\cal N}_{(D-1)a}$ 
denoted by dots turn out to be 
proportional to the {\it l.h.} sides of Eqs. (\ref{Ta}), 
(\ref{H})--(\ref{RS}), 
but {\sl not} of the Einstein equation (\ref{MD-1}) (see Sec. IV). 
Eqs. (\ref{DRS}), (\ref{DMa=})
 can be directly derived  from the definitions of 
${\Psi}$ and ${M}$ in (\ref{RS}),  
(\ref{MD-1}) (see \cite{rheo} for their complete expressions in $D$=$11$
and Appendix B for $D=4$ case). Alternatively, 
substituting 
${\delta}_{ls}$ 
in (\ref{lsusy})--(\ref{lsusyw}) or $\tilde{\delta}_{gc}$ in (\ref{lt}) 
for the generic $\delta$ in (\ref{vSSGD}) 
one finds that the coefficients of the arbitrary parameters 
$\epsilon^{\underline{\alpha}}(x)$ and $t^a(x)$ are proportional to 
${\cal N}_{(D-1)\underline{\alpha}}$ and ${\cal N}_{(D-1)a}$ 
respectively. 
Thus the local symmetries  ${\delta}_{ls}S=0$ and 
$\tilde{\delta}_{gc}S=0$ 
imply the identities  
(\ref{DRS}) and (\ref{DMa=}) 
respectively, and {\it viceversa}. 
This may be also 
easily verified in a complete form in $D=4$, $N=1$ supergravity, 
where (\ref{DRS}) and (\ref{DMa=}) 
are given by 
(\ref{DRS4}) and (\ref{DMa=4}) respectively in Appendix B,  
and where the variation 
(\ref{vSSGD}) has the form (\ref{vSSG4}).

 The Noether identity for the general coordinate symmetry in its  
${\delta}_{gc}$ form, where all the variations are the result of 
the change $\delta_{gc} x^\mu$, Eq. (\ref{gc}),  
should state that the variation of 
the action with respect to the coordinates $x^\mu$ 
does not produce 
an independent equation of motion. Thus this Noether identity should 
read   
\begin{eqnarray}\label{NIgc}
{\cal N}_\mu:= \delta S_{D,SG}/\delta x^\mu  \equiv 0\; .
\end{eqnarray}

To verify that we indeed have 
the identity  (\ref{NIgc}) in free supergravity, 
it is convenient to separate the generic variation 
$\delta$ in (\ref{vSSGD}) into the coordinate variation 
and the variation $\delta^\prime$ of the fields. 
Then, with the above notation, 
a generic variation of the action is split 
\cite{demo} as follows 
\begin{eqnarray}
\label{vSSGD1}
\delta S_{D,SG} & = \;   \int_{M^D} 
(d^Dx {\cal N}_\mu \delta x^\mu - 
\qquad {} \qquad {} \qquad \nonumber \\ & -
\int_{M^D} (-)^D 
M_{(D-1)a}\wedge dx^\mu \delta^\prime e_\mu^a(x) + \ldots \; 
\end{eqnarray}
where the dots indicate  terms containing the variations $\delta^\prime$
of the gravitino,  
spin connection fields {\it etc}.

As the action is written in terms of differential 
forms, the generic variation (\ref{vSSGD}) 
is expressed in terms of  the generic variations of these forms,  
$\delta e^a$ {\it etc}. One can split the generic 
variation of each form  into the effect produced by
the coordinate variation $\delta x$ 
(which is given by the Lie derivative, {\it e.g.} $e^a(x+\delta x)-e^a(x)
= {\cal D}(\delta x^\mu e_\mu^a) + e^b \delta x^\mu e_\mu^cT_{cb}{}^a$), 
and the field variation $\delta^\prime$. 
With such a splitting Eq. (\ref{vSSGD}) reads
\begin{eqnarray}
\label{vSSGD2}
& \nonumber \delta S_{D,SG}=   - (-)^D \int_{M^D}
M_{(D-1)a}\wedge ({\cal D}(\delta x^\mu e_\mu^a) + \hspace{-1cm} \\ \nonumber 
& +
e^b \delta x^\mu e_\mu^cT_{cb}{}^a+ \; dx^\mu \; \delta^\prime e_\mu^a) + 
\ldots  = \hspace{-1cm}
\\ \nonumber 
&   = (-)^D \int_{M^D}
({\cal D} M_{(D-1)a}\, e_\mu^a\,  -
e^b e_\mu^cT_{cb}{}^a + \ldots ) \delta x^\mu - \hspace{-1cm}
\\ 
& - (-)^D \int_{M^D}
M_{(D-1)a}\wedge dx^\mu \; \delta^\prime e_\mu^a + \ldots \; . \qquad 
\hspace{-1cm} \end{eqnarray}
Comparing (\ref{vSSGD1}) with (\ref{vSSGD2}) one  
finds the expression 
for  ${\cal N}_\mu:= \delta S_{D,SG}/\delta x^\mu$,  
\begin{eqnarray}
\label{dxN=DM} 
& d^D x \; {\cal N}_\mu =
 (-)^D ({\cal D}M_{(D-1)\; a} \; + \ldots) \; e_\mu^a \; ,  
\end{eqnarray}
where the meaning of the dots is the same as in Eqs. (\ref{DRS}), 
(\ref{DMa=}). Then the desired  Noether identity ${\cal N}_\mu\equiv 0$ 
(Eq. (\ref{NIgc})) 
for the general coordinate symmetry 
$\delta_{gc}$ in (\ref{gc}) follows immediately from 
Eq. (\ref{DMa=}) (Eq. (\ref{DMa=4}) in Appendix B for 
$D=4$, $N=1$ supergravity). 

Since $\tilde{\delta}_{gc}$ and $\delta_{gc}$ are equivalent forms of 
the same spacetime general coordinate transformations, 
it may seem strange to have two independent Noether identities 
expressing the same local invariance, Eqs. (\ref{DMa=}) and 
(\ref{NIgc}). 
However, one realizes that Eq. (\ref{dxN=DM}), written as
\begin{eqnarray}
\label{NIx}
& d^D x \; {\cal N}_\mu 
- (-)^D ({\cal D}M_{(D-1)\; a} \; + \ldots) \; e_\mu^a \;   
\; \equiv 0\; ,  
\end{eqnarray}
is also another Noether identity which reflects diffeomorphism invariance. 
Indeed,  
it implies that a combination of the equations of motion 
$\delta S_{D,SG}/\delta x^\mu\,$,  $\delta S/\delta e_\mu^a (x)\,$, {\it etc.}
and their derivatives, namely (see (\ref{DMcomp})) 
\begin{eqnarray}
\label{NIx1}
& {\delta S_{D,SG} \over \delta x^\mu} + 
\left({\cal D}_\nu {\delta S_{D,SG}\over e_\nu^a (x)}+ \ldots 
\right) \; e_\mu^a(x)\; \equiv 0\; ,  
\end{eqnarray}
vanishes identically, and thus it may replace  
either (\ref{NIgc}) or (\ref{DMa=}) as an independent identity. 
The Noether identity (\ref{NIx}) reflects 
the independence of any differential form
on the choice of the local coordinates used to write it. 
This {\sl diffeomorphism} 
invariance means that  any variation $\delta_{diff}$ of coordinates 
\begin{eqnarray}\label{dif}
& {x}^\mu \mapsto {x}^{\mu\prime}= {x}^\mu + \delta_{diff} {x}^\mu \; 
\qquad \delta_{diff} {x}^\mu & =  b^\mu ({x}) \;  
 \end{eqnarray}
can be supplemented 
by the corresponding variation of the 
component functions of the differential forms,  {\it e.g.} 
by a change of frame 
$e_\mu^a(x)
\rightarrow e_\mu^{a\prime}(x)= e_\mu^a(x) - (L_b e^a)_\mu$ 
({\it cf.} Eq. (\ref{lt})), 
\begin{eqnarray}\label{edif1}
& \delta^\prime_{diff}e_\mu^a(x)= - (L_be^a(x))_\mu \; ,
 \end{eqnarray}
in such a way 
that 
\begin{eqnarray}\label{edif}
\delta_{diff} e^a := e^{a\prime}(x^{\prime}) - e^{a}(x)=0
\;  .
\end{eqnarray}
The variation of the action vanishes trivially for such transformations,  
$\delta_{diff}S_{D,SG}\equiv 0$ since 
$\delta_{diff}$ of any differential form vanishes. 
On the other hand, splitting $\delta_{diff}S_{D,SG}$ as  
in  Eq.  (\ref{vSSGD1}), one finds that 
$\delta_{diff}S_{D,SG} =  \int_{M^D} 
(d^D x \; {\cal N}_\mu - (-)^D
{\cal D}M_{(D-1)\; a} \; e_\mu^a + \ldots ) \; b^\mu (x)$. 
Hence diffeomorphism symmetry and the Noether identity 
(\ref{NIx}) imply each other. 

This also makes explicit why the symmetry  
$\tilde{\delta}_{gc}$ (\ref{lt}) may be identified as the 
variational copy of the general coordinate transformations 
${\delta}_{gc}$ (\ref{gc}): one can obtain $\tilde{\delta}_{gc}$ 
for the parameter $t^a(x)=t^\mu(x) e_\mu^a(x)$
as a combination of 
a general coordinate transformation ${\delta}_{gc}$ with parameter 
$t^\mu$ and a change of local coordinates 
${\delta}_{diff}$
with parameter $b^\mu = -t^\mu$: 
$\;\tilde{\delta}_{gc}(t)= {\delta}_{gc}(t) + {\delta}_{diff}(b=-t)$. 
Similarly, the Noether identity 
 ${\cal N}_{(D-1)a}$ for $\tilde{\delta}_{gc}$ is 
proportional to the 
difference between the one  
for ${\delta}_{diff}$, Eq. (\ref{NIx}), 
and $dx^D {\cal N}_{(D-1)\mu} \, e_\mu^a$ (Eq. (\ref{NIgc})) for 
${\delta}_{gc}$.

\section{Equations for the coupled supergravity---bosonic brane 
system}
 
The variation of the coupled action (\ref{SSG+p}) with respect to the 
bosonic vielbein 
$\delta e^a$ produces  
the {\sl Einstein equation with 
source},  
\begin{eqnarray}\label{Ei=JT} 
& {M}_{(D-1){ {a}}} = J^{(p)}_{(D-1)a} \; ,
\\ \label{JTp=}
& J^{(p)}_{(D-1)a}\equiv (dx)^{\wedge (D-1)}_\mu 
\int_{W^{p+1}} *\hat{e}_a 
\wedge d\hat{x}^{\mu} 
\delta^D (x -\hat{x})\, ,  
\\ \label{Ei=JT0} 
& J^{(p=0)}_{(D-1)a} \equiv 
 (dx)^{\wedge (D-1)}_\mu 
\int_{W^1} d\hat{x}^{\mu} l(\tau) \hat{e}_{\tau a} 
\, 
\delta^D (x -\hat{x})\, . 
\end{eqnarray} 

The ($p$+$1$)--form gauge field equation (\ref{G}) 
evidently acquires a source from the {\sl p--brane} 
`Wess--Zumino term' $\hat{C}_{p+1}$ (\ref{LpST})
provided the gauge field $C_q$ with $q$=$p+1$ enters the 
supergravity multiplet
\begin{eqnarray}\label{Geq=J} 
& c \; G_{D-p-1} = j_{D-p-1}\; , \\ \label{JG1} 
& j_{D-p-1} = (dx)^{\wedge (D-p-1)}_{\mu_1\ldots \mu_{p+1}} 
{(-1)^{p+1}\over (p+1)!} j^{\mu_1\ldots \mu_{p+1}} \; ,\\ \label{JG2} 
& j^{\mu_1\ldots \mu_{p+1}} = 
\int_{W^{p+1}} d\hat{x}^{\mu_1}\wedge \ldots \wedge  d\hat{x}^{\mu_{p+1}} 
\delta^D(x-\hat{x}) \; . 
\end{eqnarray} 
As $w^{ab}$, $F_{a_1\ldots a_q}$ and  
the gravitino 1--form 
${\hat e}^{\underline{\alpha}} ={\hat e}^{\underline{a}} 
{\psi}_{\underline{a}}^{\underline{\alpha}}(\hat{x})$
do not appear in the bosonic $p$-brane 
action,  eqs. (\ref{Ta}), (\ref{H}) as well as 
the Rarita--Schwinger 
equation 
(\ref{RS}) 
do not acquire a source term from $\delta S_{D,p,0}$. 
Nevertheless, this does not mean that the gravitino is free.
Indeed, the covariant derivative in Eq. (\ref{RS}) involves the 
spin connection which is defined through a vielbein that satisfies the 
Einstein equation with source (\ref{Ei=JT}). 

\section{Local symmetries of  supergravity interacting with a bosonic brane}

\subsection{Local supersymmetry}

The supergravity part of the coupled system is of course invariant under 
the local supersymmetry (\ref{lsusy})--(\ref{lsusyw}) 
due to the identity (\ref{DRS}) 
(now no longer a Noether identity due to the source in Eq.
(\ref{Ei=JT})).  
Thus, to study the invariance  of the 
coupled action (\ref{SSG+p}) under $\delta_{ls}$ it is sufficient to look at 
the variation of its bosonic $p$--brane part, $\delta_{ls}S_{D,p,0}$.

Let us consider first the case of a {\sl bosonic massless particle},  $p=0$.  
Then (see Eq. (\ref{L1})),  
\begin{eqnarray}\label{dlsS}
& \delta_{ls} S_{D,0,0}=  
\int_{W^1} (l(\tau) \hat{e}_{\tau a}\,  \delta_{ls}\hat{e}^a + 
  d\tau {1\over 2} 
\hat{e}_{\tau a}\hat{e}_{\tau}^a \, \delta_{ls} l(\tau)) =
\nonumber \\ 
& = \int_{W^1} (-2i l(\tau) \hat{e}_{\tau a} 
\hat{e}^{\underline{\alpha}}
{\Gamma}^a_{\underline{\alpha}\underline{\epsilon}} 
\epsilon^{\underline{\epsilon}}
 +  d\tau {1\over 2} 
\hat{e}_{\tau a} \hat{e}_{\tau}^a\, \delta_{ls} l(\tau))\; ,   
\end{eqnarray}
so that $\delta_{ls} S_{D,0,0}=0$ {\it iff}, on $W^1$,  
\begin{eqnarray}\label{E1/2} 
& \epsilon^{\underline{\epsilon}}(\hat{x})
= \hat{e}_\tau^a
{\Gamma}_a^{\underline{\epsilon}\underline{\kappa}}
{\kappa}_{\underline{\kappa}}
(\tau)\equiv \tilde{\Gamma}^{\underline{\epsilon}\underline{\kappa}}
{\kappa}_{\underline{\kappa}}(\tau)
\; , \\  \label{vlE1/2}
& \hbox{and} \qquad  \delta_{ls}l(\tau)=4i \hat{e}_\tau^{\underline{\alpha}}
{\kappa}_{\underline{\alpha}}(\tau)\; ,
\end{eqnarray} 
where ${\kappa}_{\underline{\alpha}}(\tau)$ is a fermionic spinor function on 
the worldline; $\delta_{ls} S_{D,0,0}=0$ follows from the fact that, 
algebraically,  $\tilde{\Gamma}^{\underline{\epsilon}\underline{\kappa}}
\tilde{\Gamma}_{\underline{\kappa}\underline{\delta}}=
\delta_{\underline{\delta}}{}^{\underline{\epsilon}} 
\hat{e}_{\tau a} \hat{e}_{\tau}^a$. 

On the mass shell, where  
$\delta S /\delta l(\tau)=0$ implies 
$\hat{e}_{\tau a} \hat{e}_{\tau}^a=0$, we find that the parameter 
$\epsilon^{\underline{\epsilon}}(\hat{x})$ defined by (\ref{E1/2}) 
contains only $n/2$ nonzero components. 
Thus, $1/2$ of the local supersymmetry is broken 
on the worldline $W^1$.

It is worth stressing that 
the local supersymmetry
acts on the pull--back of the {\sl bosonic} vielbein 
$\hat{e}^a\equiv 
d\hat{x}^{\mu}(\tau) e^a_\mu (\hat{x})$ 
in the same way as 
the fermionic $\kappa$--symmetry transformation 
$\delta_\kappa$ of the {\it super}particle 
\cite{ALS} in a supergravity background, 
\begin{eqnarray}\label{Ssp} 
& S_{D,0} = 
\int_{W^1} {1 \over 2} l(\tau ) \hat{E}^a 
  \hat{E}^b_{\tau}\eta_{ab}\; ; \qquad  
\\ 
\nonumber
& \delta_\kappa S_{D,0} =0 \qquad \hbox{for}  
\\ \label{kappaZ}
& \delta_\kappa \hat{x}^{\mu} := \hat{E}_\tau^a
{\Gamma}_a^{\underline{\alpha}\underline{\epsilon}}
{\kappa}_{\underline{\epsilon}}(\tau)E_{\underline{\alpha}}^\mu 
(\hat{x}, \hat{\theta})  
\; ,
\\ \label{kappatheta}
& \delta_\kappa \hat{\theta}^{\epsilon} 
:= \hat{E}_\tau^a
{\Gamma}_a^{\underline{\alpha}\underline{\epsilon}}
{\kappa}_{\underline{\epsilon}}(\tau) 
E_{\underline{\alpha}}^{\epsilon}(\hat{x}, \hat{\theta})
\; ,
\\
\label{Sspkappa} 
&\qquad \delta_{\kappa}l(\tau)=4i \hat{E}_\tau^{\underline{\alpha}}
{\kappa}_{\underline{\alpha}}(\tau)\; , \qquad 
\end{eqnarray}
acts on the pull--back of the {\it super}vielbein
\cite{th},  
\begin{eqnarray}\label{hEaS}
& \hat{E}^a=
d\hat{x}^{\mu} {E}^a_\mu (\hat{x}, \hat{\theta})+ 
d\hat{\theta}^{\alpha} {E}^a_\alpha (\hat{x}, \hat{\theta})\; , 
\\
\label{bEfE}
& \hat{E}^{\underline{\alpha}}= d\hat{x}^{\mu} 
{E}^{\underline{\alpha}}_{\mu} (\hat{x}, \hat{\theta})+ 
d\hat{\theta}^{\beta}
E_{\beta}^{\underline{\alpha}}(\hat{x}, \hat{\theta})\; ,   
\end{eqnarray} 
where $\hat{\theta}= \hat{\theta}(\xi)$. 
In other words, 
$\delta_{\kappa} \hat{E}^a(\hat{x}, \hat{\theta}=0 ) = 
\delta_{ls}\hat{e}^a$ with 
$\epsilon^{\underline{\epsilon}}(\hat{x})$ given by (\ref{E1/2}). 
Let us stress that in the usual treatment of  $\kappa$--symmetry 
\cite{BST,AETW} super--$p$--branes are considered 
 in a {\sl superfield supergravity background} (hence   
without considering a supergravity action), {\it i.e.} by having the 
supervielbeins (\ref{hEaS}), (\ref{bEfE})  restricted by 
the {\sl superspace constraints}. These are simply the equations
that follow by extending  
($e^a \mapsto E^a$, ${e}^{\underline{\alpha}}\mapsto {E}^{\underline{\alpha}}$)
Eqs. 
(\ref{Ta}), (\ref{H}) to superspace  $(x^\mu, {\theta}^{\alpha})$.

A similar situation occurs for a $p>0$ 
bosonic brane {\sl which is the bosonic `limit'   
($\hat{\theta}^{\alpha}(\xi)= 0$) 
of a superbrane} (see also Appendix C). 
For instance,  for the $D$=$11$ membrane 
Eq. (\ref{lsusyC})  reads
$ \delta_{ls} C_{3} = e^{\underline{\alpha}} \wedge 
\bar{\Gamma}^{(2)}_{\underline{\alpha}\underline{\epsilon}}
{\epsilon}^{\underline{\epsilon}}(x) \;$    
and, hence, 
\begin{eqnarray}\label{lsSM2}
 \delta_{ls} S_{11,2,0} & =  
\int_{W^3} {1\over 2} *\hat{e}_{a} \wedge \delta_{ls}\hat{e}^a 
- \delta_{ls}\hat{C}_3 = \qquad {} \qquad \hspace{-0.5cm}
\nonumber \\
 & =\int_{W^3} (-i  *\hat{e}_{a} \wedge \hat{e}^{\underline{\alpha}}
{\Gamma}^a_{\underline{\alpha}\underline{\epsilon}} 
- \hat{e}^{\underline{\alpha}}\wedge
\hat{\bar{{\Gamma}}}^{(2)}_{\underline{\alpha}\underline{\epsilon}})
{\epsilon}^{\underline{\epsilon}}(\hat{x}) \, 
\hspace{-0.5cm} \nonumber \\
 & = -i \int_{W^3} *\hat{e}_{a} \wedge \hat{e}^{\underline{\alpha}}\, 
({\Gamma}^a(I- \bar{\gamma}))_{\underline{\alpha}\underline{\epsilon}} 
\epsilon^{\underline{\epsilon}} (\hat{x})\; , \hspace{-0.5cm} \;  
\end{eqnarray}
where 
\begin{eqnarray}
\label{bgamma}
& \bar{\gamma}\equiv {i \over 3!\sqrt{|g|}}\; \epsilon^{ijk}
\hat{e}_i^a \hat{e}_j^b \hat{e}_k^c  \Gamma_{abc} \quad  
({\hbox{tr}}\bar{\gamma}=0\; , \;  \bar{\gamma}^2=I)\; , 
\end{eqnarray}
may be recognized as the $\hat{\theta}(\xi )$=$0$ value of the 
matrix 
$\bar{\gamma}^S\equiv i/(3!\sqrt{|g|})\; \epsilon^{ijk}
\hat{E}_i^a \hat{E}_j^b \hat{E}_k^c  \Gamma_{abc}\;$ 
appearing in the $\kappa$--symmetry projector 
$(I+\bar{\gamma}^S)$ 
of the {\sl super}membrane \cite{BST} (M2--brane) 
\begin{eqnarray}\label{SM2}
& S_{11,2}= \int_{W^3} 
{\hat{\cal L}}_{3}= \int_{W^3} 
{1\over 2\, 3!} * \hat{E}_{{a}} \wedge \hat{E}^{{a}} - \hat{C}_{3}
(\hat{x}, \hat{\theta})\; .
\end{eqnarray}
Thus, when a bosonic $p$--brane is the bosonic limit of a superbrane, 
the coupled system of supergravity and this {\sl bosonic} brane  
possesses on $W^{p+1}$ $1/2$ of the original local supersymmetry 
$\delta_{ls}$ (\ref{lsusyx})--(\ref{lsusyw})
with a `$\kappa$--like' parameter 
\begin{eqnarray}\label{E1/2p}
& p> 0\; : \quad \epsilon^{\underline{\alpha}}(\hat{x})= 
(I+ \bar{\gamma})^{\underline{\alpha}\underline{\epsilon}} 
\kappa_{\underline{\epsilon}}(\xi)\; 
\end{eqnarray}
 (see (\ref{E1/2}) for $p$=$0$ where 
$(I+ \bar{\gamma}) \rightarrow 
\tilde{\Gamma}$). If the bosonic brane action is not obtained 
by setting $\hat{\theta}(\xi)=0$ in the action of a superbrane, 
then $\delta_{ls}S_{brane}=0$ would imply 
$\hat{\epsilon}^{\underline{\alpha}}(\hat{x})=0$ rather than 
(\ref{E1/2p}) since the projector $(I- \bar{\gamma})$ in (\ref{lsSM2})
would be replaced by a non--singular matrix in general. 

Out of the worldvolume, {\it i.e.} on $M^D$ but not on $W^{p+1}$,  the local 
supersymmetry is preserved completely.

\subsection{General coordinate transformations} 

In the same manner one  finds that in the coupled system  
the (variational copy of the) general coordinate symmetry (\ref{lt}) of pure 
(super)gravity  is partially broken and that their 
preserved part resembles 
the (variational copy of the) worldvolume general coordinate 
transformations (reparametrization symmetry) of the brane when acting 
on the pull--back of the 
differential forms. For instance, for  the $p=0$ coupled system 
one finds that under (\ref{lt}) $\tilde{\delta}_{gc} S =
\tilde{\delta}_{gc} S_{D,0,0}$ 
becomes ({\it cf.} Eq. (\ref{dlsS})) 
\begin{eqnarray}\label{dltS}
\tilde{\delta}_{gc} S_{D,0,0} & =  
\int_{W^1} (l(\tau) \hat{e}_{\tau a} \tilde{\delta}_{gc}\hat{e}^a + 
  d\tau {1\over 2} 
\hat{e}_{\tau a}\hat{e}_{\tau}^a  \tilde{\delta}_{gc} l(\tau)) 
\nonumber \\ 
& 
= 
\int_{W^1} (l(\tau) \hat{e}_{\tau a}\, {\cal D}t^a 
+ l(\tau) \hat{e}^b_{\tau}\, t^c T_{cb}{}^a) 
+ 
\nonumber \\ 
& \quad +  
  \int_{W^1} d\tau {1\over 2} 
\hat{e}_{\tau a}\hat{e}_{\tau}^a \, \tilde{\delta}_{gc} l(\tau)) \; ,   
\end{eqnarray}
Thus  $\tilde{\delta}_{gc} S =0$ requires that 
on $W^1\;$ $\; \hat{t}^a\;$ is expressed in terms of a single function 
$k(\tau)$ 
\begin{eqnarray}\label{lt1/2} 
& \hat{t}^a \equiv t^a (\hat{x})= \hat{e}_\tau^a k(\tau)\; , 
\end{eqnarray} 
and 
\begin{eqnarray}\label{ltl1/2} 
\tilde{\delta}_{gc} l(\tau)= - l(\tau){\cal D}_\tau k(\tau) + 
k(\tau){\cal D}_\tau l(\tau)\; .
 \end{eqnarray} 

In the general $p\geq 0$ case, 
the $D$--dimensional general coordinate symmetry 
on $M^D$, 
$\tilde{\delta}_{gc}$ in Eq. (\ref{lt}), 
is partially broken on $W^{p+1}$ down to a $(p+1)$--dimensional 
invariance on $W^{p+1}$ with 
 \begin{eqnarray}\label{ltp1/2}
\hat{t}^a \equiv t^a (\hat{x})= \hat{e}_i^a k^i(\tau)\; . 
\end{eqnarray}
Note that Eq. (\ref{lt}) with (\ref{ltp1/2}) implies
\begin{eqnarray}\label{ltWp+1}
\tilde{\delta}_{gc} \hat{e}_i^a = {\cal D}_i (\hat{e}_j^a k^j(\tau))
 + e_i^b \; e_j^c k^j \; T_{cb}{}^a
 \; , \quad  etc. \; , 
\end{eqnarray} 
which is identical to ${\delta}_{gc}\hat{e}_i^a$ produced by 
${\delta}_{gc}{x}^\mu$ with 
${\delta}_{gc}\hat{x}^\mu = k^i(\xi) \partial_i  \hat{x}^\mu(\xi)$ 
on $W^{p+1}$ (general coordinate transformations on the worldvolume).

\section{Selfconsistency conditions 
for the coupled system and fermionic equations for 
the bosonic particle}

The {\sl superbrane} fermionic equations of motion 
for {\it e.g.}, 
$p$=$0$  (see (\ref{hEaS}), (\ref{bEfE}) for notation),  
\begin{eqnarray}\label{fEqm}
& 
\hat{E}^{\underline{\epsilon}}\, 
\Gamma^a_{\underline{\alpha}\underline{\epsilon}}\, \hat{E}_{\tau a} =0 \; ,
\end{eqnarray}
have a well defined $\hat{\theta}(\xi)\rightarrow 0\;$ limit,  
$\hat{e}^{\underline{\epsilon}}\, 
\Gamma^a_{\underline{\alpha}\underline{\epsilon}}\, \hat{e}_{\tau a} =0 \;$. 
One may ask whether this limit is reproduced by the 
 dynamical system
under consideration.  
We show now that this is the case; this will turn out to be 
important in the next section to check the consistency of 
the interaction.

The particle equations of motion 
$(\delta S / \delta \hat{x}^\mu)$ $e_\mu^a(\hat{x})$= $0$
and $\delta S / \delta l(\tau)= 0$ are of course purely bosonic
\begin{eqnarray}\label{pb}
{\cal D}(l(\tau) \, \hat{e}_{\tau a}) + l(\tau)\,  \hat{e}_{\tau b}
\hat{e}^c T_{c a}{}^b(\hat{x})=0\; , 
\\ 
\label{pe} \hat{{e}}^{a}_{\tau } \hat{e}_{\tau a} =0\; .
\end{eqnarray} 
But besides Eqs. (\ref{pb}), (\ref{pe}), 
there is a nontrivial equation on $W^1$  
produced by the selfconsistency  
condition for the equation of motion  (\ref{RS}), 
\begin{eqnarray}
\label{DRS=0}
& {\cal D} {\Psi}_{(D-1)\underline{\alpha}} = 0 \; . 
\end{eqnarray}

When $p$=$0$ and the supergravity multiplet does not contain a vector field
(as, {\it e.g.}, in $D=3, 4, 11$), the gauge field equation 
(\ref{G}), the Rarita--Schwinger one 
(\ref{RS}) and the geometric equations (\ref{Ta}), (\ref{H}) 
remain sourceless; 
the source appears only in (\ref{Ei=JT}) and, thus, 
the terms denoted by dots in 
(\ref{DRS}) (and in (\ref{DMa=})) 
vanish on shell. In the interacting case these are no longer 
Noether identies
since the Einstein equation acquires a source term. But they are still 
identities algebraically satisfied by the explicit expressions of the terms 
appearing in their {\it l.h.} sides. Then, using (\ref{DRS}), 
the selfconsistency condition 
 (\ref{DRS=0}) becomes  
${\cal D} {\Psi}_{(D-1)\underline{\alpha}} = 
2i {M}_{(D-1) a}\wedge {e}^{\underline{\epsilon}}
\Gamma^a_{\underline{\alpha}\underline{\epsilon}} =0 \;$  which,   
in the light of  (\ref{Ei=JT}),  implies that the particle current  
(\ref{Ei=JT0}) satisfies the additional equation   
\begin{eqnarray} 
\label{curJ}
J^{(p=0)}_{(D-1) a}\wedge e^{\underline{\beta}}
\Gamma^a_{\underline{\alpha}\underline{\beta}} =0 \; .
\end{eqnarray}
Using the properties of the delta function one finds that  
Eq. (\ref{curJ}) is equivalent to 
$\int_{W^{1}} 
\, d\tau l(\tau) \hat{e}^{\underline{\epsilon}}_\tau \,  \hat{e}_{\tau a}
\Gamma^a_{\underline{\epsilon}\underline{\alpha}} 
\delta^D (x -\hat{x}) = 0\;$. 
After integration with a probe function this in turn implies 
the {\sl fermionic} equation on $W^1$ 
\begin{eqnarray}\label{fEqm0}
& \hat{e}^{\underline{\beta}}\, 
\Gamma^a_{\underline{\beta}\underline{\alpha}}\, \hat{e}_{\tau a} =0 \;  , 
\end{eqnarray} 
which coincides with the result of setting 
$\hat{\theta}(\tau)=0= d\hat{\theta}(\tau)$ 
in the fermionic equations of motion (\ref{fEqm})
for a massless  {\sl superparticle} (\ref{Ssp}) 
moving in a {\sl superfield supergravity 
background} (and, in particular, in flat superspace).

Similarly, one can find that the (bosonic) selfconsistency condition 
for Eq. (\ref{Ei=JT}),  
\begin{eqnarray}
\label{D(M-J)=0}
& {\cal D} ({M}_{(D-1) a}- {J}^{(p=0)}_{(D-1) a})=0\; ,
\end{eqnarray} 
being considered together with the identity (\ref{DMa=}) (no longer a   
{\it Noether} identity) 
implies the current conservation 
\begin{eqnarray}
\label{DJ=0}
{\cal D} {J}^{(p=0)}_{(D-1) a}=0\; , 
\end{eqnarray}
which 
produces the additional bosonic equation on $W^1$ 
\begin{eqnarray}
\label{D(le)}
{\cal D}(l(\tau ) \hat{e}_{\tau a})=0 \; . 
\end{eqnarray}
However, Eq. (\ref{D(le)})  becomes equivalent to Eq. (\ref{pb}) after 
Eq. (\ref{fEqm0}) is taken into account. Indeed, Eq. (\ref{Ta})
(which is not changed in the coupled system) implies (see (\ref{sgmultf}))  
$T_{cb}{}^a= - i \psi_c^{\underline{\alpha}} 
\Gamma^a_{\underline{\alpha}\underline{\beta}}
\psi_b^{\underline{\beta}}$. Thus 
the second term in Eq. (\ref{pb}), 
$\hat{e}_{\tau b} \hat{e}_{\tau}^c T_{ca}{}^b= 
- i \hat{e}_{\tau}^c \psi_c^{\underline{\alpha}} 
\Gamma^b_{\underline{\alpha}\underline{\beta}}\hat{e}_{\tau b}
\; \psi_a^{\underline{\beta}} \equiv 
- i \hat{e}_{\tau}^{\underline{\alpha}} 
\Gamma^b_{\underline{\alpha}\underline{\beta}}\hat{e}_{\tau b}
\; \psi_a^{\underline{\beta}}$, is the product of the {\it l.h.s} of 
Eq. (\ref{fEqm0}) and the pull--back of the gravitino field. 
It vanishes  when Eq. (\ref{fEqm0}) holds, and Eq. (\ref{pb}) 
becomes  (\ref{D(le)}).

\section{Gravitino interaction and its consistency}

It is well known that locally supersymmetric theories allow for a 
consistent interaction of the spin $3/2$ fields, and that this remains true 
in supergravity theories with broken local supersymmetry \cite{DZ} 
({\it e.g.} super--Higgs effect \cite{VS}). We need to check  
whether the present breaking on $W^{p+1}$ of $1/2$ of the local 
supersymmetry of free supergravity does not result 
in an inconsistency for the interacting system of supergravity 
and the bosonic brane. 

In  `free' supergravity the on--shell unwanted degrees of freedom of 
the gravitino field are removed by means of the local supersymmetry 
(\ref{lsusyf}), 
\begin{eqnarray}\label{lspsi}
& \delta_{ls}{\psi}_{\mu}^{\underline{\alpha}}= 
{\cal D}_{\mu}\epsilon^{\underline{\alpha}}+ \ldots 
\; , 
\end{eqnarray}
where the 
dots denote terms where the parameter  
$\epsilon^{\underline{\alpha}}$ enters without derivatives, 
but in a product with fields. 
Thus, in the weak field approximation, Eq. (\ref{lspsi}) reduces to 
\begin{eqnarray}\label{lspsi1}
& \delta_{ls}{\psi}_{\mu}^{\underline{\alpha}}= 
\partial_{\mu}\epsilon^{\underline{\alpha}}
\;  \quad (\hbox{weak  field  approximation})\; , 
\end{eqnarray}
obviously a gauge symmetry of the standard massless free  
Rarita--Schwinger equation 
\begin{eqnarray}\label{RSfree}
\Gamma^{abc}_{\underline{\alpha}\underline{\beta}}
\partial_b \psi_c^{\underline{\beta}}=0\quad 
(\hbox{weak  field  approximation})\; .
\end{eqnarray}

The reduction of the degrees of freedom  of 
${\psi}_{\mu}^{\underline{\alpha}}$ can be done either by using the above 
local supersymmetry in a `covariant' manner to fix  the gauge 
$\Gamma^a _{\underline{\alpha}\underline{\beta}} 
\psi_a^{\underline{\beta}}=0$ 
(which then produces $\partial^a \psi_a^{\underline{\alpha}}=0$),  
or in a noncovariant way by fixing first a Coulomb--like gauge 
$\Gamma^I_{\underline{\alpha}\underline{\beta}} \psi_I^{\underline{\beta}}=0$ 
(with {\it e.g.}, $I=1, \ldots ,(D-1)$) 
and then using the residual gauge invariance of this equation 
(that exists for $\epsilon^{\underline{\alpha}}$ satisfying 
$\Gamma^I_{\underline{\alpha}\underline{\beta}} 
\partial_I\epsilon^{\underline{\beta}}=0$) to fix 
$\psi_0^{\underline{\alpha}}=0$.  This can be made because using the free  
Rarita--Schwinger equation ({\it i.e.}, the weak field approximation) 
one finds that in the above gauge $\psi_0^{\underline{\alpha}}$ satisfies 
$\Gamma^I_{\underline{\alpha}\underline{\beta}} 
\partial_I \psi_0^{\underline{\beta}}=0$. 

The splitting 
$\psi_a^{\underline{\beta}}=
(\psi_0^{\underline{\beta}}, \psi_I^{\underline{\beta}})$ is obviously 
arbitrary and, in particular, one can replace $\psi_0^{\underline{\beta}}$
by $\hat{e}_\tau^a \psi_a^{\underline{\beta}}(\hat{x}) = 
\hat{\psi}_\tau^{\underline{\beta}}$ when the pull back  
$\hat{\psi}_a^{\underline{\beta}}$ on $W^1$ is considered. Now, 
since in the coupled system $\epsilon^{\underline{\alpha}}(\hat{x})$ is 
restricted on $W^1$ by Eq. (\ref{E1/2}) (or by 
$\epsilon^{\underline{\alpha}}(\hat{x})= 
(I+ \bar{\gamma})^{\underline{\alpha}}{}_{\underline{\epsilon}} 
\kappa^{\underline{\epsilon}}(\xi)$ 
in the $p>0$ case) we have to check that there is still enough freedom 
left in $\epsilon^{\underline{\alpha}}(x)$ to use 
Eq. (\ref{lspsi}) (Eq. (\ref{lspsi1})) 
as in the free case. 
The key point is to notice 
that Eq. (\ref{E1/2}) on $W^1$ (as well as its $p>0$ counterparts)  
does not restrict the pull--backs 
of the derivatives of the local supersymmetry parameter 
$\epsilon^{\underline{\alpha}}$ in the directions 
`orthogonal' to the worldline (worldvolume). 
In other words, among 
$({\cal D}_\mu \epsilon^{\underline{\alpha}})(\hat{x})$ 
only the combination 
$\partial_\tau\hat{x}^\mu  
({\cal D}_\mu  \epsilon^{\underline{\alpha}})(\hat{x})= 
{\cal D}_\tau \hat{\epsilon}^{\underline{\alpha}}
= {\cal D}_\tau (\hat{e}_\tau^a
\Gamma_a^{\underline{\alpha}\underline{\kappa}} 
\kappa_{\underline{\kappa}})$ is  restricted on shell by 
the arguments following (\ref{E1/2}). 
Thus the only combination of the on--shell gravitino field components 
${\psi}_{\mu}^{\underline{\alpha}}$ whose transformation rule 
(\ref{lspsi}) is restricted on $W^1$ 
is 
$\hat{\psi}_{\tau}^{\underline{\alpha}}
\equiv 
\hat{e}_{\tau}^a {\psi}_{a}^{\underline{\alpha}}(\hat{x})
\equiv 
\partial_{\tau}\hat{x}^\mu {\psi}_{\mu}^{\underline{\alpha}}(\hat{x})
$ for which the leading term is 
\begin{eqnarray}
\label{lsW}
\delta_{ls}\hat{\psi}_{\tau}^{\underline{\alpha}}= 
{\cal D}_{\tau}\hat{\epsilon}^{\underline{\alpha}}+ \ldots =
\hat{e}_{\tau}^a 
\Gamma_a^{\underline{\alpha}\underline{\kappa}} 
{\cal D}_{\tau}\kappa_{\underline{\kappa}} + \ldots 
\;   
\end{eqnarray}
since $\delta_{ls} \partial_{\tau}x^\mu {\psi}_{\mu}^{\underline{\alpha}}
(\hat{x})= 
\partial_{\tau}x^\mu \delta_{ls} {\psi}_{\mu}^{\underline{\alpha}}
(\hat{x})$ by Eq. (\ref{lsusyx}). 
However, $\hat{\psi}_{\tau}^{\underline{\alpha}}$ 
is subjected to 
Eq. (\ref{fEqm0}) which, 
since 
$\hat{e}^{\underline{\alpha}}\equiv d\tau 
\partial_{\tau}x^\mu {\psi}_{\mu}^{\underline{\alpha}}(\hat{x})\equiv d\tau 
\hat{\psi}_{\tau}^{\underline{\alpha}}$ (Eq. (\ref{sgmultf})), 
can be written as 
\begin{eqnarray}\label{fEqm1}
& \hat{\psi}_{\tau}^{\underline{\alpha}}
\Gamma^a_{\underline{\alpha}\underline{\beta}}\, \hat{e}_{\tau a} =0 \;  . 
\end{eqnarray} 
The general solution of Eq. (\ref{fEqm1}),  
$\hat{\psi}_{\tau}^{\underline{\alpha}}= \hat{e}_{\tau}^a
\Gamma_a^{\underline{\alpha}\underline{\kappa}}\, 
\nu_{\underline{\kappa}}(\tau)$,  can be gauged away, 
$\hat{\psi}_{\tau}^{\underline{\alpha}}=0$, by taking 
${\cal D}_\tau \kappa_\kappa = - \nu_{\underline{\kappa}}(\tau) + \ldots$.

Hence, despite that there is less supersymmetry on $W^1$, 
$\;{\psi}_{\mu}^{\underline{\alpha}}$ does not 
get additional degrees of freedom  on $W^1$ 
and the spin $3/2$ field still has only transversal ($2$ in $D=4$)  
polarizations in the coupled case: 
the fermionic equation (\ref{fEqm1}) replaces the broken part of  
local supersymmetry in the role of removing the unwanted gravitino degrees of 
freedom on $W^1$. As a result, 
the gravitino interaction in the supergravity---bosonic 
brane system is consistent {\it iff} $1/2$ of the original local 
supersymmetry is preserved on $W^1$.  
As we have seen in Sec. IV, this happens when the 
bosonic brane is the pure bosonic limit of a superbrane.

\section{On the 
matching of the fermionic and 
bosonic degrees of freedom}

We have concluded in Sec. IVA that the action for the  
supergravity---bosonic brane coupled system 
preserves $1/2$ of the local supersymmetry $\delta_{ls}$ of the `free' 
supergravity action. 
A faithful linear realization of 
supersymmetry requires equal number 
of bosonic and fermionic degrees of freedom. In this respect 
the above results could look surprising 
because they imply that  
local supersymmetry is still 
preserved (albeit partially) after adding 
a pure bosonic dynamical system, the $p$--brane, which   
presumably would introduce 
bosonic degrees of freedom on the worldvolume $W^{p+1}$. 

A simple way of  solving this paradox is to note that 
we are considering supergravity in the component formulation, where  
local supersymmetry acts on fields, but not on the spacetime 
coordinates (Eqs. (\ref{lsusy})--(\ref{lsusyw}), (\ref{lsusyx})). 
Thus, the brane coordinate functions 
$\hat{x}^\mu(\xi)$ 
are inert under local supersymmetry as well. 
In other words, $\hat{x}^\mu(\xi)$ are singlets  
(as, {\it e.g.} the heterotic fermions or chiral bosons in the heterotic 
superstring model), do not enter into the supergravity multiplet,  
and cannot produce a mismatch in the numbers of bosonic and fermionic degrees 
of freedom of such supermultiplet.

\section{Diffeomorphism invariance and $p$--brane 
degrees of freedom}

Nevertheless, we may go beyond the above discussion 
by showing that in the supergravity--bosonic 
brane coupled system the bosonic $p$--brane  does not carry 
any degrees of freeedom {\it i.e.},  that in the coupled system the 
$p$--brane 
degrees of freedom can be regarded as pure gauge. 
The key point is that the coupled action,
for which the original general coordinate invariance of free supergravity 
is broken on $W^{p+1}$ (Eq. (\ref{ltp1/2})), is 
constructed in terms of differential forms 
and thus it is still {\sl spacetime} diffeomorphism invariant, 
Eq. (\ref{edif}). 
This follows if the change of coordinates (\ref{dif}) is accompanied 
by 
\begin{eqnarray}\label{difW} 
\delta_{diff} \hat{x}^\mu= b^\mu (\hat{x}) \, .  
\end{eqnarray}
for the $p$--brane coordinate functions $\hat{x}^\mu(\xi)$; 
then, on $W^{p+1}$, $\hat{e}^{a\prime} (\hat{x}^\prime)= 
\hat{e}^{a} (\hat{x})$. 
This local symmetry 
suggests that  $(D-(p+1))$ brane degrees of freedom in $\hat{x}^\mu(\xi)$ 
may be regarded as  `pure gauge'. 
Indeed, whatever {\it e.g.} the massless particle worldline is, 
we can always use a general coordinate transformation to define 
(at least locally)  a  coordinate frame 
in which the worldline is represented by 
a (light--like) straight line. 
Similarly, for any worldvolume $W^{p+1}$ it is 
possible to define local coordinates on $M^D$ 
in such a way that $W^{p+1}$ is described by 
$\hat{x}^{\mu \prime}(\xi)= (\xi^i, 0, \ldots, 0)$ 
in the new coordinate system.

To justify that the bosonic diffeomorphism symmetry,  
Eqs. (\ref{dif})--(\ref{edif}) and (\ref{difW}), is indeed a gauge symmetry 
of the coupled system (but {\sl not} a {\sl gauge} symmetry for the brane 
in (super)gravity background) one can use the second Noether
theorem (see Sec. IIC). This implies the existence of a 
Noether Identity (NI) relating the {\it l.h.s.} of the equation 
$\delta S /\delta \hat{x}^{\mu}(\xi)=0$ (Eq. (\ref{pb}) for $p=0$) 
with the {\it l.h.} sides of the {\sl field} equations of the coupled 
system, $\delta S /\delta e_{\mu}^a(x)=0$ (Eq. (\ref{Ei=JT})), 
$\delta S /\delta \psi_{\mu}^{\underline{\alpha}}(x)=0$ 
(Eq. (\ref{RS})), {\it etc.}
The existence of such NI in the supergravity---bosonic brane coupled 
system has been actually proved at the end of Sec. V, where we have shown that 
the selfconsistency condition Eq. (\ref{D(M-J)=0}) for Eq. (\ref{Ei=JT}) 
produces Eq. (\ref{D(le)}) which, in turn, coincides with Eq. 
(\ref{pb}), after the selfconsistency condition  (\ref{DRS=0}) for 
Eq. (\ref{RS}) is taken into account. 
(Note that this NI is absent for the brane in a (super)gravity 
{\sl background}, because such an approximation cannot produce 
the Einstein equation with a dynamical, {\it i.e.} 
$\hat{x}^{\mu}(\xi)$--dependent, source.) 

Thus, we may conclude that {\sl the bosonic brane does not 
carry any degrees  of freedom in the coupled system} described by the 
sum of (super)gravity action and the brane action 
(see Appendix D for further discussion).

\section{Final remarks} 

The above suggests that the fermionic degrees of freedom of 
the superbrane in the supergravity---superbrane interacting 
system might be considered as pure gauge as well {\it i.e.}, 
that the superbrane degrees of freedom coupled to 
dynamical ({\it i.e.} not background) supergravity 
are pure gauge ones. Thus, one would expect that 
in the (singular) `gauge' $\hat{\theta}(\xi)=0$ any model for  
supergravity interacting with a dynamical superbrane source  
produces an action closely related  
(or equivalent) to (\ref{SSG+p}). 
Indeed, the coupled system (\ref{SSG+p}) is the result of `fixing' 
the singular gauge $\hat{\theta}^\alpha (\xi)$=$0$ 
in the supergravity---super--$p$--brane coupled system described by the 
sum of the group manifold action for 
supergravity and the {\sl superbrane} action \cite{BAIL}.

This result is not so surprising as it might look at first sight.
The possibility of gauging away the superstring degrees of freedom just 
reflects the fact that, by an appropriate choice of the 
coordinates, the brane may be located arbitraryly with respect to 
a coordinate system. This does not mean, however, 
that the coupled system is gauge equivalent 
to `free' supergravity, since the source term in the Einstein equation 
(\ref{Ei=JT}) cannot be removed by a gauge transformation,
although the freedom to choose arbitrary 
local coordinates may simplify 
the expression for the current (Eqs. (\ref{JTp=}), (\ref{Ei=JT0})).

This gauge character of the brane degrees of freedom in the presence of 
{\sl dynamical} supergravity ({\it i.e.}, described by an action 
rather than being introduced as a 
background) might be looked at as a property `dual' to the fact that 
(linearized) supergravity appears  in the quantum states spectrum of 
a superbrane (superstring) in {\sl flat} superspace, although it is not 
present independently.

We conclude by the following observation. If we considered our  
supergravity---bosonic $p$--brane  system with $p=D-2$ on the orbifold 
spacetime $M^D=M^{D-1}\times [S^1/\hbox{\bf Z}_2]$ in an approximation 
where the $(D-2)$--brane is fixed at the orbifold fixed `points', 
the $(D-1)$--hyperplanes,  and the brane dynamics were ignored, we would 
arrive at a model of the type considered in refs. \cite{BKPO,Bergshoeff}. 
Then the observed explicit breaking of $1/2$ of the 
local supersymmetry of `free' supergravity by the bosonic brane
would correspond to the vanishing, at the orbifold 
fixed `points',  of the part of the supersymmetry 
parameter which is odd under the $\hbox{\bf Z}_2$ projection,  
a characteristic of the models \cite{BKPO,Bergshoeff}. 
However, our analysis indicates that the (partial) supersymmetry preservation 
is not a specific property of  models on the orbifold spacetime  
$M^D=M^{D-1}\times [S^1/\hbox{\bf Z}_2]$ but rather that 
it is inherent of any Lagrangian description of the 
supergravity---bosonic $p$--brane 
interacting system, for any $p$, 
provided 
(see Sec. IVA) that the $p$--brane is the 
$\hat{\theta}(\xi)=0$ `limit' of a super--$p$--brane.

\medskip

{\it Acknowledgments}. 
The authors are grateful to Dmitri Sorokin and Mario Tonin 
for useful discussions at the early stages of this work. 
This work has been partially supported by 
the DGICYT research grant PB 96-0756,  
Ucrainian FFR  (research project $\# 383$)  
and the Junta de Castilla y Le\'on (research grant C02/199). 
Two of us (I.B. and J.A.) thank the 
Erwin Schr\"{o}dinger Institute for Mathematical Physics 
and the organizers of the {\sl Mathematical Aspects of String Theory} 
program for their kind hospitality during the final stages of this work. 

\bigskip 

\renewcommand{\theequation}{A.\arabic{equation}} 
\setcounter{equation}0

\section*{Appendix A:} 

\subsection*{On the formulation of spacetime general coordinate 
transformations}

Eq. (\ref{gce1}) reflects the action of a general coordinate transformation 
$x^\mu \rightarrow x^{\mu \prime}= x^\mu + t^\mu(x)$ on the vielbein one--form 
$e^a(x)=dx^\mu e_\mu^a(x)$; it is given by the Lie derivative 
$L_t e^a(x) := e^a(x^\prime)-e^a(x)
\equiv e^a(x+t) -e^a(x)
= dx^\nu (t^\mu \partial_\mu e_\nu^a +
\partial_\nu t^\mu \;e_\mu^a)\,$.
Note that since differential forms are invariant under 
diffeomorphisms {\it i.e.}, they can be expressed in any local 
coordinate system ($e^{a\prime}(x^{\prime})=e^a(x)$, 
{\it etc.}, where the primes refer to the new coordinate 
system), the Lie derivative may 
also be expressed as $- L_t e^a(x) := e^{a\prime}(x) - e^a(x)$. 
In fact, $e^{a\prime}(x^\prime )- e^a(x)=
e^{a\prime}(x^\prime )-e^{a}(x^\prime )+ e^a(x^\prime )- e^a(x)=
\delta^\prime_{diff} e^a(x) +  L_b e^a(x)=
-L_b e^a(x^\prime ) + L_b e^a(x)=0$ since 
$L_b e^a(x^\prime )= L_be^a(x)$ at first order. 

Using that 
$L_t e^a(x)= (di_t + i_td)e^a(x)
= dt^a + i_t (de^a)$ 
we find 
\begin{eqnarray}
\label{gc00}
& {\delta}_{gc}e^a(x) = L_t e^a(x)
= 
{\cal D}t^a(x) + i_t T^a + e^b i_t w_b{}^{a}\; , 
\end{eqnarray}
where $t^a:= i_te^a = t^\mu e_\mu^a$,  
$i_t T^a= e^b t^c T_{cb}{}^a$ and the last term  
$i_t w_b{}^{a}= t^c w_{cb}{}^{a}$ is a local Lorentz transformation induced 
by the local translations $t^\mu(x)$. 
Introducing the `covariant' Lie derivative,  
\begin{equation}
\label{covLd}
{\cal L}_t:= i_t{\cal D} + {\cal D}i_t \; , 
\end{equation}
 we see that 
${\cal L}_t e^a= {\cal D}i_te^a + i_t{\cal D}e^a$, where 
${\cal D}e^a = T^a$. Thus 
${\delta}_{gc}e^a(x)$ in (\ref{gc00}) is the sum of 
${\cal L}_t e^a$ and of the induced Lorentz rotation $e^b i_t w_b{}^{a}$, 
${\delta}_{gc}e^a(x)={\cal L}_t e^a+ e^b i_t w_b{}^{a}$.

For the full  supergravity multiplet components we have 
\begin{eqnarray}
& {\delta}_{gc}e^a(x) & := L_t e^a(x) = 
{\cal L}_t e^a(x) + e^b i_t w_b{}^{a}(x)=   
 \nonumber \\ \label{gc001} 
&& = {\cal D}t^a + i_t T^a + e^b i_t w_b{}^{a}\; , 
\\ \nonumber
& {\delta}_{gc}e^{\underline{\alpha}}(x) & 
:= e^{\underline{\alpha}}(x+t)- e^{\underline{\alpha}}(x)
= L_t e^{\underline{\alpha}}(x)
= \\ 
 & & = 
{\cal L}_te^{\underline{\alpha}}(x) + 
e^{\underline{\beta}} i_t w_{\underline{\beta}}{}^{\underline{\alpha}}(x)
\; , \label{gcf}
\\ \nonumber 
& {\delta}_{gc} C_q(x) & := C_q(x+t)-C_q(x) = L_tC_q(x)= 
\\ \label{gcC00}
&& = i_tdC_q + d(i_t C_q)  
\; , 
\\ \label{gcw00} 
& {\delta}_{gc}w^{ab}(x) & := L_t w^{ab}(x) 
= i_tR^{ab} + {\cal D}(i_tw^{ab})\; ,  
 \\ 
\nonumber 
& {\delta}_{gc}F^{a_1\ldots a_{q+1}}&(x)  := 
L_t F^{a_1\ldots a_{q+1}}(x) = \\ \nonumber 
&& 
= t^\mu \partial_\mu 
F^{a_1\ldots a_{q+1}}  (x)= \\ \nonumber 
&& = 
i_t {\cal D}F^{a_1\ldots a_{q+1}}(x) + \\ 
\label{gcF} 
&& \qquad + 
(q+1)  
F^{[a_1\ldots a_{q}|b} i_tw_b{}^{|a_{q+1}]}(x)  \, , 
\end{eqnarray}
where 
$i_te^{\underline{\alpha}}(x):=t^\mu (x)\psi_\mu^{\underline{\alpha}}(x)$, 
$i_tw_{b}{}^a(x):=t^\mu w_{\mu b}{}^a(x)$ and 
$i_tw_{\underline{\beta}}{}^{\underline{\alpha}}(x) := 
t^\mu w_{\mu\underline{\beta}}{}^{\underline{\alpha}}(x)= 
(1/4) i_tw^{ab}\Gamma_{ab}{}_{\underline{\beta}}{}^{\underline{\alpha}}$, 
$\; {\cal D}t^a=dt^a - t^b w_b{}^a$, 
${\cal D}e^{\underline{\alpha}}(x)= de^{\underline{\alpha}}(x)- 
e^{\underline{\beta}} 
w_{\underline{\beta}}{}^{\underline{\alpha}}(x)$, 
${\cal D}(i_t w^{ab})= d(i_t w^{ab}) - (i_t w^{ac}) \; w_c{}^{b}
+ (i_t w^{bc}) \; w_c{}^{a}$ and 
${\cal D}F^{a_1\ldots a_{q+1}}(x)= 
dF^{a_1\ldots a_{q+1}}(x)- (q+1)  
F^{[a_1\ldots a_{q}|b}(x) w_b{}^{|a_{q+1}]}(x)$.

In a theory with manifest local Lorentz symmetry the $i_t w_b{}^{a}$ 
terms in ${\delta}_{gc}e^a$, ${\delta}_{gc}e^{\underline{\alpha}}$, 
${\delta}_{gc}w^{ab}$, ${\delta}_{gc}F^{a_1\ldots a_{q+1}}$ 
may be conveniently ignored, as well as the 
$d(i_t C_q)$ term in a theory with the gauge invariance 
$\delta_{gauge}C_q = d\Lambda_{q-1}$. In this case 
the general coordinate variations of the fields,  
Eqs. (\ref{gc001})--(\ref{gcF}), reduce to    
\begin{eqnarray}
\label{gc0010}
& {\delta}_{gc}e^a(x) = {\cal D}t^a(x) + i_t T^a \; , 
\\ 
\label{gcf1}
& {\delta}_{gc}e^{\underline{\alpha}}(x) 
= {\cal D}i_te^{\underline{\alpha}}(x) + 
i_t {\cal D}e^{\underline{\alpha}} \; , 
\\ \label{gcC001}
& {\delta}_{gc} C_q = i_tdC_q   \; , \qquad  
\\ \label{gcw001}
& {\delta}_{gc}w^{ab} = i_tR^{ab} \; , \qquad   
 \\ \label{gcF1} 
& {\delta}_{gc}F^{a_1\ldots a_{q+1}}(x) = 
i_t {\cal D}F^{a_1\ldots a_{q+1}}(x) \; .
\end{eqnarray}
The $D=4$, $N=1$ superspace counterparts of these transformations 
are called `supergauge transformations' in \cite{BaggerW}.

The component functions $(L_t e^a(x))_\mu$, say,  
of the 
one--form $L_t e^a(x)$  are given by 
$(L_te^a(x))_\mu= t^\nu \partial_\nu e_\mu^a +
\partial_\mu t^\nu \, e_\nu^a \,$. Since 
$L_t(dx^\mu e_\mu^a(x))= L_t(dx^\mu)\;  e_\mu^a(x) + 
dx^\mu L_t e_\mu^a(x)$, 
we see that the term $\partial_\mu t^\nu \, e_\nu^a$ 
in $(L_te^a(x))_\mu$ 
comes from 
$L_t(dx^\mu)= d(L_t x^\mu)= d(\delta x^\mu)= dt^\mu$. 
Thus, if we now {\it define} the variation of the component 
function $\tilde{\delta}_{gc}e_\mu^a(x)$ by 
$(L_te^a)_\mu(x)$ above,  
\begin{eqnarray}
\label{gcvc}
& \tilde{\delta}_{gc}e_\mu^a(x) 
= {\cal D}_\mu t^a + e_\mu^b t^c T_{cb}{}^a \; , 
\end{eqnarray}
then there is no need to vary $x^\mu$ since 
the effect of its variation has been already taken into account in  
(\ref{gcvc}). Thus we may set $\tilde{\delta}_{gc}x^\mu=0$ and 
use $\tilde{\delta}_{gc}$ given by Eq. (\ref{gcvc})  (Eq.  (\ref{lt}))
as an equivalent definition of the general coordinate 
transformations ${\delta}_{gc}$ when varying an action 
constructed from differential forms. 
For the other supergravity fields $\tilde{\delta}_{gc}$ is given by 
\begin{eqnarray}
\label{gcvcf1}
&  \tilde{\delta}_{gc}e_\mu^{\underline{\alpha}}(x) 
= {\cal D}_\mu i_te^{\underline{\alpha}}(x) + 
e_\mu^b t^c ({\cal D}e^{\underline{\alpha}})_{cb} \; ,  
\\ \label{gcvcw}
&  \tilde{\delta}_{gc}w_\mu^{ab} = e_\mu^c t^d R_{dc}{}^{ab} \; , \qquad   
 \\ \label{gcvcF} 
&  \tilde{\delta}_{gc}F^{a_1\ldots a_{q+1}}(x) = 
t^\mu {\cal D}_\mu F^{a_1\ldots a_{q+1}}(x) \; ,
\end{eqnarray}
to which one may add
\begin{eqnarray}
\label{gcvcC}
&  \tilde{\delta}_{gc} C_{\mu_1 \ldots \mu_q}  
= (q+1) t^{\nu} \partial_{[\nu }  C_{\mu_1 \ldots \mu_q]}
  \; , \qquad  
\end{eqnarray}
ignoring the induced gauge transformations $d(i_tC_q)$. 

The variation  
$\tilde{\delta}_{gc}$ is the so--called `variational version' 
\cite{WZ} of the general coordinate transformations ${\delta}_{gc}$
that act on forms 
as in Eq. (\ref{gce1}).

\renewcommand{\theequation}{B.\arabic{equation}} 
\setcounter{equation}0

\section*{Appendix B: }
\subsection*{
Noether identities for $D=4$, $N=1$  supergravity}

For $D=4$ simple ($N=1$) supergravity, where the supergravity 
action contains only the  two terms (\ref{LD2}), (\ref{LD3/2}), 
\begin{eqnarray}
\label{SSG4}
& S_{4,SG}= \int_{M^4} 
{1\over 2} \epsilon_{abcd} R^{ {a} {b}} \wedge e^c \wedge e^d \; 
+ \nonumber 
\\ & \qquad + {2i\over 3} \int_{M^4} 
\epsilon_{ {a} {b} {c} d} 
{\cal D} e^{\underline{\alpha}} \wedge e^{\underline{\beta}}
\wedge 
e^{a}\; 
{\Gamma}^{bcd}_{\underline{\alpha}\underline{\beta}} \; .
\end{eqnarray}
The complete expression for the 
Noether identities (\ref{DRS}), (\ref{DMa=}) is  
\begin{eqnarray}\label{DRS4}
& {\cal N}_{4\underline{\alpha}}\equiv &
{\cal D} {\Psi}_{3\underline{\alpha}} - 
2i {M}_{3 a}\wedge {e}^{\underline{\beta}}
\Gamma^a_{\underline{\alpha}\underline{\beta}} - 
\\ \nonumber & & - \Gamma_{a\underline{\alpha}\underline{\beta}}\; 
{\cal D}e^{\underline{\beta}}\wedge 
(T^a + i e^{\underline{\gamma}}\wedge e^{\underline{\delta}} 
\Gamma^a_{\underline{\gamma}\underline{\delta}})\equiv 0\; ,
\\ 
\label{DMa=4} 
& {\cal N}_{4a}:= & {\cal D} {M}_{3 a} - \qquad {} \qquad \\ \nonumber 
& & - {1\over 2} \epsilon_{abcd} R^{bc} \wedge 
(T^d + i e^{\underline{\alpha}}\wedge e^{\underline{\beta}} 
\Gamma^d_{\underline{\alpha}\underline{\beta}})\equiv 0\; .
\end{eqnarray}
The the {\sl generic} variation of the action (\ref{SSG4}) is 
({\it cf.} (\ref{vSSGD}))
\begin{eqnarray}
\label{vSSG4}
& \delta S_{4,SG}= - \int_{M^4}
M_{3a}\wedge \delta e^a  - \int_{M^4} \Psi_{\underline{\alpha}} \wedge 
\delta e^{\underline{\alpha}} \\ \nonumber 
& + {1\over 3!} \int_{M^4} \epsilon_{abcd} e^a \wedge 
(T^b + i e^{\underline{\gamma}}\wedge e^{\underline{\delta}} 
\Gamma^b_{\underline{\gamma}\underline{\delta}}) \wedge \delta w^{cd}\; .
\end{eqnarray}

\renewcommand{\theequation}{C.\arabic{equation}} 
\setcounter{equation}0

\section*{Appendix C: }
\subsection*{Bosonic brane in the Polyakov--like formulation
and local supersymmetry }

The discussion in Sect. VIA of the $p=0$ case involved the einbein as an 
initially independent field (see (\ref{L1})). The analogue in the $p>0$ cases
requires treating also independently the worldvolume metric. 
This may be done by using the 
Brink--Di Vecchia--Howe--Polyakov \cite{P} formulation,  
\begin{eqnarray}\label{LpST0}
{\hat{\cal L}}^\prime_{p+1}=  
{1\over 4} * \hat{e}_{{a}} \wedge \hat{e}^{{a}} 
- {(p-1)\over 4} (-)^p *1 - \hat{C}_{p+1}\; , 
\end{eqnarray}
where the Hodge star operator $*$ involves the auxiliary worldvolume metric
 $g_{ij}(\xi )$,  $\; 
* \hat{e}_{{a}} \wedge \hat{e}^{ {a}}= d^{p+1}\xi \sqrt{|g|}g^{ij}  
\hat{e}^{a}_i \hat{e}^b_{j} \eta_{ab}$ and
$\; (-)^p *1$= $d^{p+1}\xi$ $\sqrt{|g|}$.  
The metric $g_{ij}(\xi)$ is determined by its equations of motion,    
\begin{eqnarray}\label{indm}
{\delta S^\prime_{D,p,0} \over \delta g^{ij}(\xi)}=0 \quad \Rightarrow \quad 
g_{ij} = \hat{e}^{a}_i \hat{e}^b_{j} \eta_{ab}\; . 
\end{eqnarray}
The $\delta_{ls}$ variation of $S$, with 
$S_{D,2,0}$ replaced by $S^\prime_{D,2,0}$ for (\ref{LpST0}),  
contains, in addition to (\ref{lsSM2}) (where now the $*$ is defined with an 
independent $g_{ij}(\xi)$), a term proportional to 
\begin{eqnarray}\label{indmv} 
& \Delta_{ls}g^{ij} 
(g_{ij}-  \hat{e}^{a}_i \hat{e}_{aj})\; , 
\end{eqnarray}
where 
\begin{eqnarray}\label{indmv1}
& \Delta_{ls}g^{ij}\equiv  \delta_{ls}g^{ij}
-{1\over 2} \delta_{ls}g^{kl} g_{kl} g^{ij} - \qquad \\ \nonumber & \qquad  
- {2\over \sqrt{|g|}}
\hat{e}_k^{\underline{\alpha}}
(\Gamma_{bc})_{\underline{\alpha}\underline{\beta}}
\hat{\epsilon}^{\underline{\beta}} \, g^{k(i}\epsilon^{j)pq}
\hat{e}_p^b\hat{e}_q^c \, , 
\end{eqnarray}
 and $\delta_{ls}g^{ij}$ is to be 
determined from $\delta_{ls}S^\prime_{D,2,0}=0$.  
The matrix $\bar{\gamma}\equiv {i\over 3!\sqrt{|g|}} \epsilon^{ijk}
\hat{e}_i^a \hat{e}_j^b \hat{e}_k^c  \Gamma_{abc}\;$ has now the properties 
$tr(\bar{\gamma})=0\;$, $\; \bar{\gamma}^2=\, 
({\sqrt{|det (\hat{e}^a_i\hat{e}_{aj})|}/\sqrt{|g|}}) \, I\;$, 
where $g := {\hbox{det}} g_{ij}$. On the mass shell 
${\sqrt{det (\hat{e}^a_i\hat{e}_{aj})}=  \sqrt{|g|}}$ due to Eq. (\ref{indm}), 
$\; \bar{\gamma}^2=I\,$ and $(I- \bar{\gamma})$ is a projector. 
Hence, again $\delta_{ls}S^\prime_{D,2,0}=0$ for 
${\epsilon}^{\underline{\epsilon}}(\hat{x})= 
(I+ \bar{\gamma})^{\underline{\alpha}\underline{\beta}} 
\kappa_{\underline{\beta}}(\xi)\;$, for which we 
can define $\delta_{ls}g_{ij}(\xi)$ in such a way that it compensates 
the contributions from the $\delta_{ls}\hat{e}^{a}$ and 
$\delta_{ls}\hat{C}_{p+1}$ variations.

Thus we  conclude that the $n$ parametric local supersymmetry is
partially broken on $W^{p+1}$ down to its
$n/2$ parametric subgroup.  

The above reasonings are 
in correspondence with the proof of the $\kappa$--symmetry 
of the supermembrane \cite{BST}.

\medskip 

\section*{Appendix D:}

\subsection*{Graviton degrees of freedom in the 
(super)gravity---$p$-brane coupled system}

In Sec. VIII we have shown that, since the coupled   
(super)gravity---bosonic $p$--brane system 
possesses diffeomorphism invariance, Eqs. (\ref{dif})--(\ref{edif}) with 
(\ref{difW}), the brane degrees of freedom are pure gauge. 

One could ask, nevertherless, whether after fixing the  gauge 
$\hat{x}^{\mu}(\xi)= (\xi^i, 0, \ldots, 0)$ the brane degrees of 
freedom could reappear somehow in the vielbein, in the sense that one 
would not have enough gauge freedom left to reduce the vielbein 
degrees of freedom on the worldvolume to the required ${(D-2)(D-1)\over 2}-1\;$ 
($=2$ for $D=4$) transversal ones. However, the degrees of 
freedom of the graviton $e^a_\mu(x)$ can be fixed by 
using general coordinate invariance, Eq. (\ref{gc}) or  (\ref{lt}).
Then, since the general coordinate transformations 
symmetry is partially broken on $W^{p+1}$ in the coupled system, the 
question is whether there is enough general coordinate symmetry 
left to fix the degrees of freedom in $e^a_\mu$ on $W^{p+1}$.

An analysis similar to that presented above for the gravitino 
indicates that this is indeed possible. 
In `free' supergravity, after removing from 
$D^2$  the ${D \choose 2 }$ antisymmetric 
degrees of freedom by 
local Lorentz invariance, the remaining
unwanted $2D$ degrees of freedom of the vielbein field 
can be disposed of by 
making use of the  general coordinate symmetry 
$\tilde{\delta}_{gc}$, Eq. (\ref{lt}). 
We now show that this can be done 
in the present supergravity--bosonic brane coupled 
system, despite that 
the components of the vector parameter $t^a$ of $\tilde{\delta}_{gc}$,  
 arbitrary out of $W^{p+1}$, are reduced to $(p+1)$ independent 
functions on $W^{p+1}$ 
(see Eqs. (\ref{lt1/2}),  (\ref{ltp1/2})).

To remove the (on--shell) unwanted $2D$ degrees of freedom 
of $e_\mu^a$ mentioned above,  one uses 
the derivatives of $t^a$
\begin{eqnarray}\label{lt1}
& \tilde{\delta}_{gc} e_\mu^a = {\cal D}_\mu t^a +\ldots \; , 
 \end{eqnarray}
({\it cf.} (\ref{lspsi})). 
This can be achieved fixing the gauge covariantly,  
$\partial^\mu  e_\mu^a =0$  (assuming the weak field approximation) and 
then using the residual `d'Alembertian' gauge invariance 
($D+D$ conditions), or by fixing a Columb--like gauge 
$\partial^I  e_I^a =0$ for the `orthogonal' part and 
and then using its residual `harmonic' gauge invariance to fix the 
`tangential' part of the gauge $\partial^0  e_0^a =0$ 
(again, $2D$ conditions). 

As before, the separation 
$(e_0^a, e_I^a)$ of components in $e_\mu^a$ 
is arbitrary. Our task now is to show that, in the 
interacting case, Eq. (\ref{lt1}) can still be used to fix a counterpart 
of these 
gauges. 
The `tangential' component of the 
derivative $({\cal D}_\mu t^a) (\hat{x})\; $ 
along $W^1\;$, 
$\;(\partial_\tau \hat{x}^\mu 
{\cal D}_\mu t^a) (\hat{x})= {\cal D}_\tau \hat{t}^a=  
{\cal D}_\tau( \hat{e}_\tau^a k(\tau))$, 
depends on a single function $k(\tau)$ on $W^1$, 
while the 
`orthogonal' components of $({\cal D}_\mu t^a) (\hat{x})$ 
are unrestricted  on $W^1$. 
Again, the only component of $e_\mu^a$ whose transformation rule 
(\ref{lt1}) is restricted on $W^1$ is 
the pull--back $\hat{e}_\tau^a= \partial_\tau \hat{x}^\mu e_\mu^a(\hat{x})$
 for which   
\begin{eqnarray}\label{lte1}
& \tilde{\delta}_{gc} \hat{e}_\tau^a= {\cal D}_\tau \hat{t}^a+ \ldots= 
\hat{e}_\tau^a\; {\cal D}_\tau k(\tau) + \ldots \; 
 \end{eqnarray}
at leading order 
({\it cf.} (\ref{lsW})) since  
$\tilde{\delta}_{gc} \partial_\tau \hat{x}^\mu e_\mu^a(\hat{x})= 
\partial_\tau \hat{x}^\mu \tilde{\delta}_{gc} e_\mu^a(\hat{x})$ 
by Eq. (\ref{lt}). 
 Thus the only danger of appearance of additional degrees of freedom 
(with respect to those in the `free' supergravity case) 
comes from the `tangential' part $\hat{e}_\tau^a$
for which,  by (\ref{lte1}), only one gauge function $k(\tau)$ is available. 
However, just this part 
is restricted on $W^1$ by Eq. 
(\ref{D(le)}), 
\begin{eqnarray}
\label{D(le)1}
{\cal D}_{\tau }(l(\tau ) \hat{e}_{\tau}^a)=0 \; 
\end{eqnarray}
({\it cf.} (\ref{fEqm1})). 
The worldline field $l(\tau )$ can be removed by the remaining freedom in 
$k(\tau)$ (more precisely, as (\ref{lte1}) has basically 
the form of infinitesimal scale transformations of $\hat{e}_{\tau}^a$ 
with the parameter $(1+ {\cal D}_{\tau }k(\tau))$, we can choose
${\cal D}_{\tau }k(\tau)$ in such a way that the 
scaling results in $\hat{e}_{\tau}^a\mapsto 
l^{-1} \hat{e}_{\tau}^a$). 
Then, Eq. 
(\ref{D(le)1}) becomes the counterpart of the gauge fixing condition 
$\partial_\tau \hat{e}_\tau^a + \ldots=0$.

\end{multicols}

\begin{thebibliography}{99}
\bibitem{demo}
At this stage we are following a field--space democracy approach  
by treating the coordinates $x^\mu$ and the fields $e_\mu^a(x)$,  
{\it etc.} on the same footing. 
This is certainly natural 
when the interacting 
dynamical system includes $p$--branes since these are described by  
{\sl coordinate} functions $\hat{x}^\mu(\xi)$ on 
$W^{p+1}$.
\end{thebibliography}
\end{document}